\documentclass[prl,superscriptaddress,twocolumn,aps,showpacs]{revtex4}
\usepackage{amssymb}
\usepackage{graphicx}
\usepackage{longtable}
\usepackage{amsmath}
\usepackage{amsfonts}
\begin{document}
\author{E.~Pavarini}
\author{E.~Koch}
\affiliation{Institut f\"ur Festk\"orperforschung and Institute for Advanced Simulations, Forschungzentrum J\"ulich, 
             52425 J\"ulich, Germany}
\author{A.I.~Lichtenstein}     
\affiliation{Institute of Theoretical Physics, University of Hamburg, 
             Jungiusstrasse 9, 20355 Hamburg, Germany}        
\date{\today }
\title{On the mechanism for orbital-ordering in KCuF$_3$}
\begin{abstract}
The Mott insulating perovskite KCuF$_3$ is considered the archetype of an orbitally-ordered system. By using the 
LDA+dynamical mean-field theory (DMFT) method, we investigate the mechanism for orbital-ordering (OO) in this material. We show that
the purely electronic Kugel-Khomskii super-exchange mechanism (KK)  alone leads to a remarkably large transition temperature of
$T_{\rm{KK}}\sim 350$~K.  However, orbital-order is experimentally believed to persist to at least 800~K. Thus
Jahn-Teller distortions are essential for stabilizing orbital-order at such high temperatures.
\end{abstract}
\pacs{71.10.Fd, 71.10.-w,71.27.+a,71.10.Hf }
\maketitle

In a seminal work \cite{kugel} Kugel and Khomskii showed that in 
strongly correlated systems
with orbital degrees of freedom 
many-body effects could give rise to orbital-order (OO) via a purely
electronic super-exchange mechanism. 
Orbital-ordering phenomena are now believed to play a crucial role in determining the electronic and magnetic properties of many transition-metal oxide Mott insulators. While it is clear that Coulomb repulsion is a key ingredient, it remains uncertain whether it just  enhances the effects of lattice distortions \cite{JT}  or
really drives orbital-order via superexchange \cite{kugel}.

We analyze these two scenarios for the archetype of an orbitally-ordered material, KCuF$_3$ \cite{kugel}. 
In this 3$d^9$ perovskite the Cu 
$d$-levels are split into completely filled three-fold degenerate $t_{2g}$-levels
 and two-fold degenerate $e_g$-levels, occupied by one hole.
In the first scenario Jahn-Teller elongations of some Cu-F bonds split the 
partially occupied $e_g$-levels further into two 
non-degenerate crystal-field orbitals. The Coulomb repulsion, $U$, then suppresses quantum orbital fluctuations favoring the occupation 
of the lower energy state, as it happens in some $t_{2g}$-perovskites \cite{evad1,evad2}.
In this picture the ordering is caused by electron-phonon coupling; Coulomb repulsion just enhances the orbital polarization due to the crystal-field splitting \cite{evad1,poteryaev}. 
In the second scenario the purely electronic super-exchange mechanism,
arising from the $e_g$-degeneracy, drives orbital-ordering,
and  Jahn-Teller distortions are merely a secondary effect. In this picture electron-phonon coupling is of minor importance \cite{kugel}. 

The key role of Coulomb repulsion is evident from static mean-field LDA+$U$ calculations, 
which show  \cite{Sasha,ldaminimum}  that in KCuF$_3$ the distortions of the octahedra
are stable with a energy gain $\Delta E\sim 150-200$ meV per formula unit,
at least an order of magnitude larger than in  LDA \cite{Sasha,ldaminimum} and GGA \cite{ldaminimum,vollhardt}; 
recent GGA+DMFT \cite{vollhardt} calculations yield very similar results,
suggesting in addition that dynamical fluctuations play a small role in determining the stable crystal structure 
of this system. 
\begin{figure}
\center
\rotatebox {0}{\includegraphics [width=0.49\textwidth]{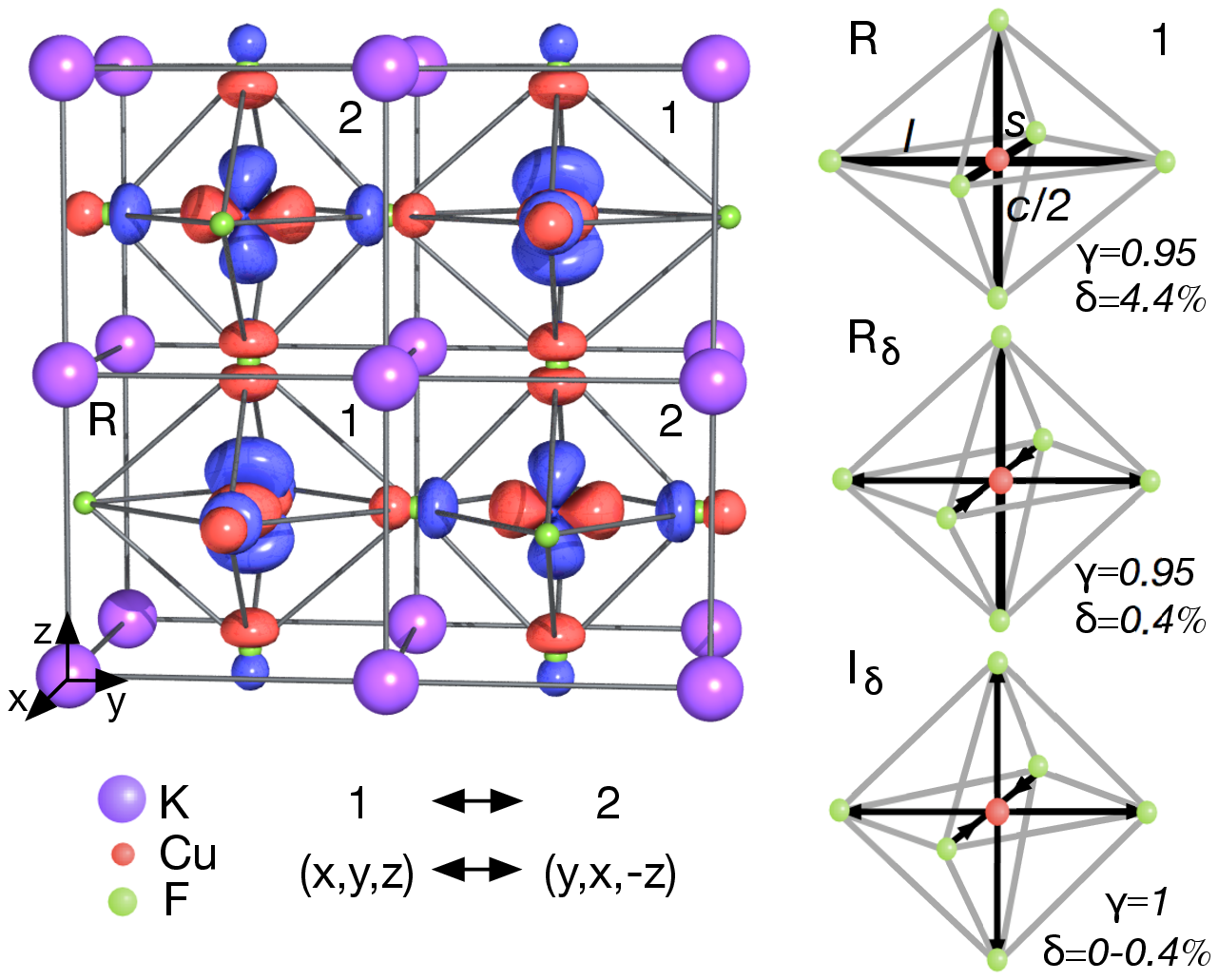}}
\caption{(Color online) \label{str} Left: Crystal structure and orbital-order in  $a$-type  \cite{poly} KCuF$_3$.  
Cu is at the center of F  octhaedra enclosed in a K cage. The conventional cell is tetragonal with axes {\bf a}, {\bf b}, {\bf c}, where $a\!=\!b$, 
$c\!=\! 0.95 \: a \sqrt2$. The pseudo-cubic axes are defined as ${\bf x}=({\bf a}+{\bf b})/2$,
${\bf y}=( -{\bf a}+{\bf b})/2$, and ${\bf z}={\bf c}/2$. 
All Cu sites  are equivalent. For sites 1 the long (short) bond $l$ ($s$) is along ${\bf y}$ (${\bf x}$). Vice versa for sites 2.
Orbital $|2\rangle$ (see table~\ref{hops}), occupied by one hole, is shown for each site.
Right:  Jahn-Teller distortions  at sites 1, measured by $\delta=(l-s)/(l+s)/2$  and $\gamma=c/a\sqrt{2}$. 
$R$  is the  experimental structure,  $R_\delta$ and $I_\delta$ two ideal structures
with reduced distortions,  and $I_0$ is  cubic.
}
\end{figure}
However, these results might merely indicate that the electron-phonon coupling is underestimated in LDA or GGA, probably due to self-interaction, rather than identifying Kugel-Khomskii 
super-exchange as the driving mechanism for orbital-order. 
This is supported by ab-initio Hartree-Fock (HF) calculations  which give results akin to LDA+$U$ \cite{HF}.
Moreover, in the superexchange scenario 
it remains to be explained why $T_{OO}\sim 800$~K  \cite{caciuffo}, more than twenty times the 3D antiferromagnetic (AFM) critical temperature, $T_N\sim38$~K \cite{hutc,poly}, a surprising fact if magnetic- and orbital-order were driven by the same super-exchange mechanism.

In this Letter we study the Kugel-Khomskii mechanism at finite temperature and identify the 
origin of orbital-ordering in KCuF$_3$. 
We  will show that  super-exchange alone leads to orbital-order with $T_{\rm{KK}} \sim 350$~K,
less than half the experimental value. Thus
Jahn-Teller distortions are  essential for driving   orbital-order above 350~K.

KCuF$_3$ is a tetragonal perovskite made of Jahn-Teller distorted CuF$_6$ octahedra 
enclosed in an almost cubic K cage \cite{str,str2}.
The Jahn-Teller distortion amounts to a 3.1$\%$ elongation/shortening   
of the CuF distances in the ${xy}$-plane.
The tetragonal distortion reduces the CuF bond along ${\bf z}$ by 2.5$\%$,
leaving it of intermediate length.
The long ($l$) and short ($s$) bond alternate between {\bf x} and {\bf y} along all three 
cubic axes ($a$-type  pattern) \cite{poly}.
At each site one hole occupies the highest $e_g$-orbital, 
$\sim |{s^2-z^2}\rangle$, i.e., the occupied orbitals 
($\sim |{x^2-z^2}\rangle$ or $ \sim |y^2-z^2\rangle$)  alternate in all 
directions. This ordering and the crystal structure  are shown in Fig. \ref{str}.

\begin{figure}
 \center
 \rotatebox {270}{\includegraphics[width=2.15in]{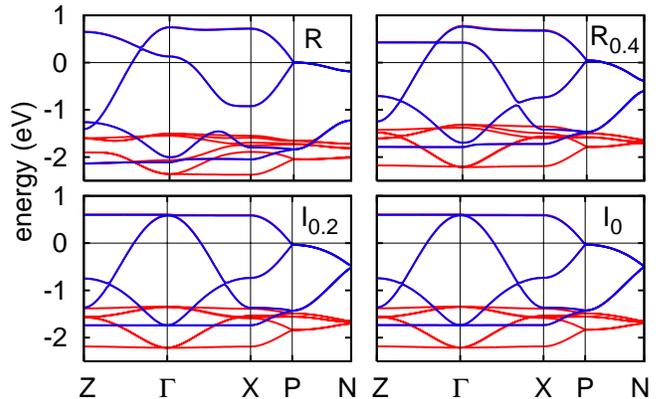}}
  \caption{(Color online) LDA 3$d$-bands (in eV) of  KCuF$_3$ for the real crystal (R) and less distorted structures (see Fig. \ref{str}). Blue: $e_g$-bands. Red: $t_{2g}$-bands. The Fermi level is at zero.}
\label{bnds}\end{figure}

As a method for studying the electronic structure of KCuF$_3$ and the super-exchange mechanism 
we adopt the LDA+DMFT approach \cite{lda+dmft}. 
Following the procedure presented in Ref.~\cite{evad1},  we first calculate the LDA band-structure using the $N^{th}$-order muffin-tin orbital method  (NMTO).  We find filled O-bands divided by a gap of $\sim 0.8$~eV from the $d$-bands (Fig.~\ref{bnds}) of
width $W_d\sim3.2$~eV ($W_{t_{2g}}\!\sim\!1$~eV, $W_{e_g}\!\sim\! 2.9$~eV).
 The energies of the $d$-crystal-field orbitals (wrt Fermi level) are  $-2.01$~eV, $-1.82$~eV, and $-1.74$~eV ($t_{2g}$) and $-1.39$~eV and $-0.34$~eV ($e_g$). 
The $t_{2g}$ states are completely filled, do not hybridize with the $e_g$-levels and thus are likely unimportant for orbital-ordering \cite{test}.
For the active states we construct a basis of localized  $e_g$ NMTO Wannier functions \cite{evad1}.
The corresponding  $e_g$ Hubbard model is

\begin{eqnarray}
 \nonumber
 H &=&H^{\mathrm{LDA}} 
   +\sum_{ im } U_{m,m} n_{im\uparrow }n_{im\downarrow}\\   &+&
   \! \frac{1}{2}\sum_{ im\left( \neq m^{\prime}\right) \sigma\sigma^\prime} 
 (U_{m,m^\prime}-J\delta_{\sigma,\sigma^\prime})n_{ im\sigma}   n_{im^{\prime}\sigma^\prime},
\label{H}
\end{eqnarray}
where $n_{im\sigma}=c_{im\sigma}^{\dagger}c^{\phantom{\dagger}}_{im\sigma}$ and $c_{im\sigma}^{\dagger}$ creates an electron with spin $\sigma$ in a Wannier orbital $|m\rangle=|x^2-y^2\rangle$ or $|3z^2-1\rangle$  at site $i;$ the direct and exchange \cite{fresard} terms of the screened on-site Coulomb interaction are $U_{mm^{\prime}}\!\!=U\!\!-2J(1-\delta_{m,m^\prime})$
and $J$.
We solve (\ref{H}) using  dynamical mean-field theory in the single-site approximation (DMFT)~\cite{DMFT} and 
its cluster extension (CDMFT) \cite{meier}, using quantum Monte Carlo \cite{hirsch} as impurity solver and working with  the full self-energy matrix $\Sigma_{mm^{\prime}}$ in orbital space \cite{evad1}. We obtain the spectral matrix on the real axis by analytical continuation \cite{jarrel}.
We use as parameters  $J=0.9$~eV and vary $U$ between $7$ and $9$~eV.
These values are close to the theoretical estimates based on constrained LDA \cite{Sasha}.

\begin{figure}
 \center
 \rotatebox{270}{\includegraphics[width=2.4in]{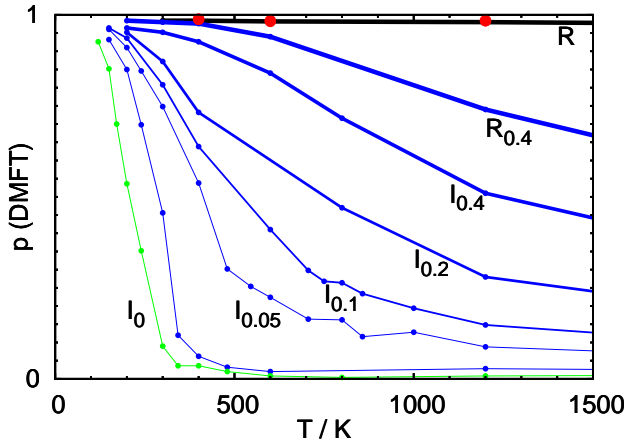}}
 \label{dmft}
 \caption{\label{pol} (Color online)
  Orbital polarization $p$  as a function of temperature calculated with LDA+DMFT 
  (black, blue, green).
  Black: $U=7$ eV, experimental structure. Blue: $U=7$ eV, idealized  structures $R_\delta$ and $I_\delta$ (see Fig.~\ref{str}) 
  with decreasing crystal-field. Green: $U=9$~eV, $I_{0}$ only.  
  Red: cluster DMFT for the experimental structure and $U=7$ eV.
  }
\end{figure}

In the paramagnetic phase, single-site DMFT calculations yield a Mott gap of about 2.5~eV for $U=7$~eV,
and 4.5 eV for $U=9$~eV. The system is orbitally ordered, and the  OO is $a$-type as the distortion pattern;
static mean-field (LDA+$U$, HF) calculation \cite{altarelli,ldaminimum,HF} give similar orbital-order,
however also antiferromagnetism. We define the orbital polarization $p$ as the difference in occupation 
between the most and least occupied natural orbital (diagonalizing the $e_g$ density-matrix).
It turns out that to a good approximation $p$ is given by the difference in occupation
between the highest ($|2\rangle$) and the the lowest ($|1\rangle$) 
energy crystal-field orbital, defined in table~\ref{hops}. 
In Fig.~\ref{pol} we show $p$ as a function of temperature.
We find that the polarization is saturated ($p\sim 1$) even for temperatures as high as 1500~K. 
We obtain very similar results in two-site CDMFT calculations.  

Using second-order perturbation theory, we calculate the exchange-coupling constants 
for the orbitally ordered state found with DMFT,  and obtain \cite{SE}
$$
J_{SE}^{i,i^\prime}\sim\frac{4|t^{i,i^\prime}_{2,2}|^2 (U+\Delta)} {(U+\Delta)^2-J^2 }-                                    
\frac{|{t^{i,i^\prime}_{1,2}|^2+|t^{i,i^\prime}_{2,1}|^2}} {U+\Delta -3J}
                   \frac{2J}{U+\Delta-J}\, ,
$$
where $t_{j,j^\prime}^{i,i^\prime}$  are the hopping integrals from site $i$ to site $i^\prime$, and $j,j^\prime=1,2$ are the $e_g$ crystal-field states. As shown in table~\ref{hops}, the calculated exchange couplings are in very good agreement with experimental findings.
Thus our method gives both the correct orbital-order and the correct magnetic structure.

\begin{table}
 \begin{tabular}{crrrrrrrrrrr}
  &\multicolumn{7}{c}{Experimental structure  $R$ \cite{str}}&\\
  $lmn$ & $t_{1,1}^{i,i^\prime}$ & $t_{1,2}^{i,i^\prime}$ & $t_{2,1}^{i,i^\prime}$ & $t_{2,2}^{i,i^\prime}$ &&$J_{SE}^{i,i^\prime} $(th.)&$J_{SE}^{i,i^\prime}$(exp.)\\ 
 \hline
  $001$  &  -83 & -161 & -161 & -343&&      57   &  53\cite{Tennant,Satija} \\
  $010$  &   98 & -352 &  -17 &   59 &&     -4 & -0.4\cite{Satija}\\
  $100$  &   98 &  -17 & -352 &   59 &&    -4 & -0.4\cite{Satija}\\ \\
  &\multicolumn{7}{c}{Ideal cubic structure $I_0$} &\\ 
   $lmn$&  $t_{1,1}^{i,i^\prime}$ & $t_{1,2}^{i,i^\prime}$ & $t_{2,1}^{i,i^\prime}$ & $t_{2,2}^{i,i^\prime}$& &$J_{SE}^{zz \;i,i^\prime}$(th.)& 
   $J_{SE}^{+- \;i,i^\prime}$(th.)
   \\ \hline   
    $001$  & -93 &   -163 &   -163 &   -282 && -162 & 0\\
    $010$  &  188 & -327& -1 &  0 &&  -40 & 60\\
     $100$  & 188 & -1 & -327&  0 &&  -40 & 60\\
    \end{tabular}
  \begin{tabular}{crrrrrrrrrrrrrr}
  \multicolumn{10}{c}{Crystal-field splittings $\Delta_{2,1}$  and $\cos\theta$ } \\
   & $R$ & $R_{0.4}$ &$R_0$&$I_{4.4}$& $I_{0.4}$ & $I_{0.2}$ & $I_{0.1}$ & $I_{0.05}$ &$I_0$ \\
  \hline
$\Delta_{2,1}$ &1050&216&180&845&89&46&22&11&0 &  \\
$\cos\theta$  & 0.99 &  0.950&  0.866&0.97 &0.97 &0.97 & 0.97& 0.97 & 0.97& 
 \\
 \end{tabular}
 \caption[]{ Hopping integrals $t_{j,j^\prime}^{i,i^\prime}$ in the crystal-field basis ($j,j^\prime$)
     from a site $i$ of type 1 to a neighboring site $i^\prime$ of type 2 in direction $l{\bf x}+n{\bf y}+m {\bf z}$.
 $J_{SE}$  are the magnetic superexchange couplings for the experimental structure and some representative 
 orbital-superexchange  couplings 
for the ideal cubic structure  \cite{SE}.
 The crystal-field states 
 are $|1\rangle=\cos\theta|3l^2-1\rangle+\sin\theta|s^2-z^2\rangle$ and
 $|2\rangle=\!-\sin\theta|3l^2-1\rangle+\cos\theta |s^2-z^2\rangle$ where
 $s$ ($l$) is the  direction of the short (long) bond ($s\!=\!x$, $l\!=\!y$ for a site 1).  
 The crystal-field splitting and  $\cos\theta$ are given for all structures.
  All energies are in meV.
  }\label{hops}
\end{table} 

To understand
whether this orbital-order is driven by the exchange coupling or merely is a consequence of the crystal-field splitting, we consider hypothetical lattices with reduced deformations, measured by $\gamma=c/\sqrt{2}a$ (tetragonal distortion) and $\delta=(l-s)/(l+s)/2$ (Jahn-Teller deformation).
To keep the volume of the unit cell at the experimental value, we renormalize all lattice vectors by $(\gamma/0.95)^{-1/3}$. We calculate the Hamiltonian for a number of structures reducing the distortion of the real crystal \cite{str} with $\gamma=0.95$ and $\delta=4.4\%$ to the ideal cubic structure $\gamma=1$ and $\delta=0$. 
The bands for some of these structures are shown in figure \ref{bnds}. 
We use the notation $R_\delta$ for structures with the real tetragonal distortion
$\gamma=0.95$ and $I_\delta$ for ideal ($\gamma=1$) structures.
The distortions affect the hopping integrals, both along (001) and in the $xy$-plane, as shown in table \ref{hops}. The main effect is, however, the crystal-field splitting $\Delta_{2,1}$ which decreases almost linearly with decreasing distortion, as expected for a Jahn-Teller system.
For each structure we obtain  the Hamiltonian $H^{\mathrm{LDA}}$ for the e$_{g}$-bands and 
perform LDA+DMFT calculations for decreasing temperatures.
At the lowest temperatures 
we find $a$-type OO with full orbital polarization for all structures (see Fig.~\ref{pol}).
At 800~K, the situation is qualitatively different. 
Orbital polarization remains saturated when reducing $\delta$ from $4.4\%$ to $1\%$. 
For smaller distortions, however, $p$ rapidly goes to zero: For $\delta=0.2 \%$ and $\gamma=1$, $p$ is already reduced 
to $\sim 0.5$, and becomes negligible in the cubic limit.
Thus super-exchange alone is not sufficiently strong to drive orbital-ordering at $T\gtrsim800$~K.

From the temperature dependence of the orbital polarization we can determine 
the transition temperature $T_{\rm{KK}}$ at which the Kugel-Khomskii superexchange mechanism  would drive
orbital-ordering, and thus disentangle the superexchange from the electron-phonon coupling. 
For this we study the ideal cubic structure, introducing a  
negligible (1 meV) crystal-field splitting as an external-field to break the symmetry.
We find a phase transition to an orbitally-ordered state at  $T_{\rm{KK}}\sim350$~K. 
The hole orbitals at two neighboring sites are $\sim|y^2-z^2\rangle$ and $\sim |x^2-z^2\rangle$, 
in agreement with the original prediction of Kugel and Khomskii \cite{kugel}.
This critical temperature is sizable, but significantly smaller than $T_{OO}\sim 800$~K \cite{caciuffo}. 

 Since the screened Coulomb repulsion $U$ is hard to calculate, 
 it is crucial to identify the range of plausible values of $U$ and 
 to estimate how $T_{\rm{KK}}$  varies in this range. 
 For the experimental structure,  $U\sim5$~eV yields a tiny gap in DMFT
 and $U\sim6$~eV a semiconducting gap of 1.3~eV, while KCuF$_3$ is a good insulator \cite{GAP}. 
 It seems therefore unrealistic that $U$ is smaller than $7$~eV. 
 It could, however, be larger. In Fig.~\ref{pol}   
 we show the results for $U=9$~eV. We find a reduction to $T_{\rm{KK}}\sim 300$~K, 
 reflecting the decrease in the super-exchange coupling. 
 Thus we can conclude that, within single-site DMFT, 
 $T_{\rm{KK}}\sim 300-350$~K, sizable but at least a factor 
 two smaller than $T_{OO}\sim 800$~K.

{ The ordering temperature $T_{\rm{KK}}$ might be even overestimated by the single-site DMFT approximation, 
as is common for mean-field theories \cite{FM}.
To investigate the effects of short-range correlations, we
therefore perform two-site CDMFT calculations for the cubic structure.
We use a supercell containing 8 formula units, with axis  
{\bf a$^\prime$}= 2{\bf x}, {\bf b$^\prime$}= 2{\bf y}, {\bf c$^\prime$}= 2{\bf z}
and  a two-site cluster which averages the cubic directions, i.e., imposing 
$\Sigma^{i,i^\prime}=\Sigma^{i,i\pm x}=\Sigma^{i,i\pm y}=\Sigma^{i,i\pm z}$, 
 for nearest neighbors $i$ and $i^\prime$ in the supercell. 
For  U=7 eV we find that the polarization starts to increase around 300-350 K,
somewhat below the single-site transition}.

To compare these results with super-exchange theory,
we  calculate the orbital super-exchange coupling assuming that no long-range magnetic order is present \cite{SE}. 
Two representative couplings are 
$$
J_{SE}^{zz \;i,i^\prime}= 
\sum_{\tau,\tau^\prime} (-1)^{\tau+\tau^\prime}\frac{2|t_{\tau,\tau^\prime}^{i,i^\prime}|^2}{U+J}\; \frac{2U+6J}{U-3J},
$$
$$
J_{SE}^{+- \;i,i^\prime}=\sum_{\tau\tau^\prime} \left[
\frac{2t_{\tau\tau}^{i,i^\prime}t_{\tau^\prime\tau^\prime}^{i,i^\prime}}{U-J}\; \frac{2U}{U-3J}
+\frac{2t_{\tau,\tau^\prime}^{i,i^\prime}t_{\tau^\prime,\tau}^{i,i^\prime}}{U-J} 
\; \frac{2J}{U+J}\right],
$$
where we adopt the pseudospin description of the orbital states \cite{kugel}, with 
$\tau=1/2$ corresponding to $|{3z^2-1}\rangle$ and $\tau=-1/2$ to $|{x^2-y^2}\rangle$.
These couplings, shown in table I, are very anisotropic, with the largest about three times the magnetic exchange coupling along {\bf {z}}. 
Again this suggests that $T_{\rm{KK}}$ should be larger than  $T_N=38$~K, but certainly smaller than T$_{OO}\sim 800~$K.

All this indicates that in KCuF$_3$ the driving mechanism for orbital-ordering is not pure superexchange.
Further support comes from an accurate re-analysis of LDA+$U$ results. To do this we first perform LDA+$U$ calculations
for the cubic and distorted structure \cite{ldaminimum}, and obtain results in agreement with previous literature \cite{Sasha,altarelli}.
For the undistorted KCuF$_3$ we find a fully orbitally polarized solution, and the occupation matrix only 
slightly differs from that obtained for the experimental structure. This shows that the energy gain due to the distortions,
$\Delta E\sim 180$ meV per formula unit, cannot be ascribed to the orbital polarization itself, but rather 
is an estimate of the electron-phonon coupling, enhanced by self-interaction  correction \cite{ldaminimum}.
In order to estimate the energy gain due to orbital polarization, we performed several LDA+$U$ calculations for the cubic
structure with different (fixed) occupation matrix. We find that the energy gain \cite{kugel}
is $\sim 90$ meV, only half of $\Delta E$.

In conclusion, 
we have calculated the Kugel-Khomskii transition temperature  for KCuF$_3$ and find a remarkably large 
$T_{\rm{KK}}\sim350$~K.  Nevertheless, the super-exchange mechanism is not sufficiently strong  
to explain T$_{OO}\sim 800$~K: at such a transition both super-exchange and electron-phonon
are of comparable importance.  The assignment T$_{OO}\sim 800$~K \cite{caciuffo} is however  based on 
the temperature dependence of the orbital peak intensity signals measured by resonant x-rays scattering \cite{caciuffo}.
A direct measurement of the evolution of the distortions with temperature would be highly desirable.
Should orbital-order persist till melting, the electron-phonon coupling contribution
would be the dominant mechanism.

We acknowledge discussions with and helpful comments from D.I.~Khomskii and P. Ghigna.
The calculations were performed on the J\"ulich BlueGene/L under account JIFF2200.


\begin{thebibliography}{99}
\bibitem{test} 
For the experimental structure
we performed LDA+DMFT calculations for the 5$d$-band Hubbard model. 
We find that the $t_{2g}$-bands are just shifted to lower energy and broadened.
The $e_g$ spectral-function shows a gap of $\sim$2.5 eV
and is almost identical to that obtained 
solving the 2-band $e_g$ Hubbard model (with $t_{2g}$ downfolded) for the same parameters.
\bibitem{SE}
We adopt the  following conventions for the superexchange couplings.  Spin-spin interaction: $\frac{1}{2}\sum_{i,j}J_{SE}^{ii^\prime}S_i\cdot S_{i^\prime}$, where $i$ and $j$ are the lattice sites.
Orbital-orbital interaction: $\frac{1}{2}\sum_{i,i^\prime,\mu,\mu^\prime}J_{SE }^{\mu \mu^\prime \; ii^\prime}T^\mu_i\cdot T^{\mu^\prime}_{i^\prime}$, where $\mu,\mu^\prime=x,y,z$ and $T$ are pseudo-spin 1/2 operators \cite{kugel}.
\end{thebibliography}
\end{document}